\documentclass[final,3p,times,twocolumn]{closerarticle}

\usepackage{graphicx}
\usepackage{amssymb,amsmath}
\usepackage{color}
\usepackage{fancyhdr}
\usepackage{gensymb}
\usepackage{amsthm}
\definecolor{red}{rgb}{1,0,0}
\definecolor{green}{rgb}{0,1,0}
\definecolor{blue}{rgb}{0,0,1}

\theoremstyle{plain}

\theoremstyle{definition}
\newtheorem{defn}{Definition}[section]

\theoremstyle{remark}

\def \titulo {On the multiplicative effect of government spending\\ (or any other spending for that matter)}%
\def \autor {Jo\~ao Pires~da~Cruz}%

\journal{Submitted to SSRN}
\begin{document}

\pagestyle{fancy}
\fancyfoot{} 
\fancyfoot[L]{Submitted to SSRN}
\fancyfoot[R]{\hspace{2.0in} \thepage}


\begin{frontmatter}

\title{\titulo}

\author{\autor$^{a,b}$}

\address{$^a$Closer Consulting Ltd, Knowledge Dock Business Centre,
University of East London, 4-6 University Way, London E16 2RD, United Kingdom}

\address{$^b$Center for Theoretical and Computational Physics, 
        			Faculdade de Ci\^encias
			Campo Grande, Edif\'icio C8
			P-1749-016 Lisboa, Portugal}
 
\address{$^{\ast}$ Phone: +44(0) 20 8223 7809; Email: joao.cruz@closer.pt}

\begin{abstract}
There is, among the economist ecosystem,  the idea of virtuous public spending as a form of promotion of economic growth. If we think on the way GDP is measured, it is not possible to get that conclusion because it becomes circular: measuring the money flow obviously will detect directly the public spending but always mixed with the flow of money from other sources. The question is how virtuous is public spending \textit{per se}?  Can it promote economic growth? Is there multiplicative effect in GDP bigger than $1$? In this paper, we make use of the first principles of Economics to show that government  spending is, at the most, as virtuous as private consumption and can be a source of economic depression and inequality if it is not restricted to fundamental services.     
\end{abstract}

\begin{keyword}
Public Economics, Public Expenditure, Quantitative Methods, General Equilibrium

\end{keyword}

\end{frontmatter}




\section{Introduction}
\label{sec:intro}

The issue of the multiplicative effect of government  spending has been a matter of discussion in the past years due to the economic crisis, both in Europe and in the USA\cite{Hall2009, Ramey2009, Christiano2009} and also because some empirical mistakes\cite{Blanchard2013} gave rise to a considerable ideological discussion which we will not approach here.

What seems interesting in the problem is that every approach made from the Economics point of view is based on econometric methods for which there is no support for fundamental assumptions, namely the invariance of the probability space. 

The causality between government spending, or austerity, and economic growth is questionable, to say the least. Specially, when we are presented with curves of a few points or modeling methods where  economic equilibrium is an assumption and not a result.

Here we approach the problem in a different perspective. First, we will ignore completely empirical data of GDP or public spending because it is not possible to establish causality between both measures. Second, we will use the conditions for which a thermodynamic equilibrium is formed to model the problem and, finally, we take conclusions based on the analytical approach. 

Why is the thermodynamic equilibrium better than the economic equilibrium for this purpose? The concept of economic equilibrium is something that is hardly\cite{Mosini2008} understood in the so called hard sciences, because it is postulated. In thermodynamics there are very strict conditions where one can say that a system is at equilibrium. But once that equilibrium is recognized, then a complete set of tools become available to understand the system. One of which is statistical physics and the invariance of the probability space. What this invariance means is that a system can be in a huge number of possible states, but that number does not change in time. This invariance is the mathematical condition for us to be able to take conclusions about causality of random variables, for example.

So, in the next section we will model an economic system, based on what economists say that an economy is. From there, we make the transformations needed to establish the conditions for a thermodynamic like equilibrium and, only then, we will take conclusions about our specific problem, which is `government spending'.

\section{Model}
\label{sec:model}

Unlike the usual Economics paper, we will not model this problem assuming that equilibrium will be formed and, from that, build a linear equation to fit empirical data. We acknowledge that this method is the usual one in Economics, but we do not want to get the same doubtful conclusions.
 
Thus, let us go through a logical process for the sake of clarity. We know that some of the definitions are already made in several ways in economics but we will define them in the scope of the model. We emphasize that reader should have in mind that whenever we refer to some concept defined bellow we will be always referring to the definition we made in the scope of this model.

\begin{figure}
   \includegraphics[width=0.9\linewidth]{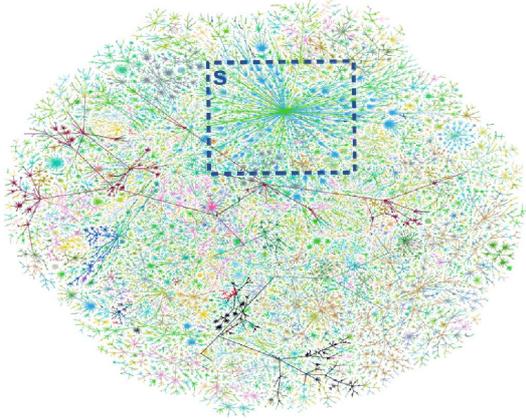}
   \caption{Scale-free network representation of an economy. A node represents an economic agent and links represent resource allocations(see text). The section S of the network represents the perimeter of a government.  (Adapted from: Hal Burch and Bill Cheswick Internet Mapping Project) }
\label{pic:scf}
\end{figure}

\begin{defn}[Economic agent]
\textit{Economic agent is an object that dissipates physical energy using resources needed by him to produce new resources needed by others}.      
\label{def:agent}
\end{defn}
And
\begin{defn}[Economic system]
\textit{Economic system is a set of economic agents that is big enough to be considered infinite for practical purposes and closed in the sense that the total number of agents is constant}.      
\label{def:system}
\end{defn}

The concept in definition \ref{def:agent} is a direct consequence of the well accepted theoretical definition of what an economy is\cite{Robbins2007} and the full explanation can be found in an introductory book\cite{Hoag2002}. The general idea is that an economic agent has two characteristics: It produces an amount of resources based on the amount of resources he can access and he exchanges those resources with others for the resources he needs. Naturally, money is one of such resources and in today's economy is present in almost every exchange. Most of the time, we will refer to economic agent simply as agent.  Definition \ref{def:system} must be done in order that we can mathematically go from the discrete domain to continuous domain and back without problems.

Thus, from definition \ref{def:agent} we will add four other definitions that we need for our model,
\begin{defn}[Wealth]
\textit{The wealth of an economic agent is the total amount of resources allocated to him}.      
\label{def:wealth}
\end{defn}

\begin{defn}[Product]
\textit{The product of an economic agent is the amount of resources added to the amount he already allocated to him as a result of its physical effort}.      
\label{def:product}
\end{defn}

\begin{defn}[Exchange]
\textit{Exchange is the co-allocation of resources between economic agents}.      
\label{def:exchange}
\end{defn}

\begin{defn}[Channel]
\textit{Channel  is a way how a set of agents exchange collectively with other agents}.      
\label{def:channel}
\end{defn}

The first three definitions are so basic in Economics that several versions of them can be found in a single bookstore. Nonetheless, and since we are following a logical process of reasoning, we have to define them. We should only make a note on definition \ref{def:product}  because it makes the bridge between the physical world and the economic world.  In the end, economy is something that derives from human effort, from physical work, that is obtained through the dissipation of chemical energy, we get economic labor. This was important for us to solve the egg-chicken problem of the existence of an economy and the circularity of the definition of resource that we did not made: If resources come from product and product from resources, what is the first resource from which the first product came from? The answer is `human effort'.

The definition \ref{def:channel} corresponds to legal concept of collective person that can be explained making use of an example. A company pays the employee for the labor to pay back shareholders. A company is not an agent in our model, because agents dissipate physical energy. A company
is a channel, a way how shareholders exchange  collectively with an employee or consumer agent or  consumer channel.

It is hardly a conundrum to say that  there is a quantity $\beta$ real, positive and finite that we can consider the average agent production in the economic system such that
\begin{equation}
\frac{dx_i}{x_i} = \beta
\label{eq:beta}
\end{equation}   
where $x_i$ is the wealth of the agent $i$, $d x_i$ the product. Eq. (\ref{eq:beta}) is just the mathematical expression of the previous definitions. To find out what this $\beta$ is all about, let us now define $\Lambda$ as the total wealth of the system. The average wealth is $\Lambda/N$ where $N$ is the number of agents in the system.  This statement is restricted to one instant in time because, since we do not know yet if the system is at equilibrium, we are only allowed to average over the set of agents 
in the system at constant time.

Let us think abstractly on the total wealth of the system, $\Lambda$.  If it was possible for the agents to produce without exchange, then the average production in the system, $\beta$, would be 
\begin{equation}
 \beta=\frac{d \left(\Lambda/N\right)}{\left(\Lambda/N\right)}=\frac{d\Lambda}{\Lambda},
 \label{eq:beta2} 
 \end{equation}
where the second equality on the r.h.s. derives from the fact that we assume $N$ constant. This means that the average relative growth in resources in equal to the total relative growth. This result is quite trivial, but the interesting characteristic of the agents is that they allocate resources to each other to produce. Their growth if \textit{correlated}. That characteristic is not embedded  in Eq.(\ref{eq:beta2}) because we are just summing wealth abstractly.  That brings us to another definition:

\begin{defn}[Gross Product(adapted from\cite{Hoag2002})]
\textit{Gross product is the total currency value of production in a subsystem}.      
\label{def:GDP}
\end{defn}

In definition \ref{def:GDP} we introduce for the first time money and it is the way GDP, $\Omega$,  is measured, by measuring the currency leg in a exchange. We are measuring only one leg of the exchange but the wealth of both counterparts is affected (one gets money and the other the product). 
So, numerically, let us think that we have a system with $\Omega=100$ dollars, and an exchange of $1$ dollar is made. Then $d\Omega/\Omega=0.01$. But $\Omega$ is just the total money leg of the exchanges made before, so $\Lambda \sim 200$  (we did not use the equal sign because $\Lambda$ is not expressed in currency units). So the average wealth is $x_i=\Lambda/N=200/N$. But in the exchange $dx_i=1/N$ because the wealth of two agents growth by $1$ not one agent by $2$. That means that $dx_i/x_i=1/2 d\Omega/\Omega$.  This is the equivalent to say that, as the GDP grows, the average wealth of the agents goes with the square root of the GDP growth. Thus, when we introduce a measuring unit, money, then the way we can express the average wealth is by 
\begin{equation}
 2 \beta=2 \frac{dx_i}{x_i} = \frac{d \Omega}{\Omega}.
 \label{eq:beta3} 
 \end{equation}
where $\Omega$ is expressed in currency units. The reader can think of the relation between wealth and gross product as the relation between the area of a circle and the radius. As the radius grows linearly, the area grows quadratically with the radius. Thinking about relative growth, the relative growth of the area is twice the relative growth of the radius.   

So, the relative growth of gross product is actually twice the average relative growth of agents' wealth.  Finally, considering the general case, we write
\begin{equation}
\frac{dx_i}{x_i} = \frac{1}{\alpha} \frac{d \Omega}{\Omega}
\label{eq:multiplicative}
\end{equation}   
where $\alpha$ is the coefficient representing correlation, $\alpha=1$ represents completely independent growth and $\alpha=2$ represents a completely correlated growth. The reader should not understand the coefficient $\alpha$ as the usually called `correlation coefficient between two random variables', represented by $\rho$ or $corr(X,Y)$. That coefficient $\rho$ is a specific measure of correlation called the Pearson coefficient, which we do not use here.  

What happened between Eq.(\ref{eq:beta2}) and Eq.(\ref{eq:multiplicative})? The measured overall quantity is not the same. While $\Lambda$ is an abstract quantity representing wealth, $\Omega$ is expressed in currency units representing the total amount of money used in the exchanges. Which allows us to express wealth in currency units, also, because Eq.(\ref{eq:multiplicative}) is dimensionless, despite the fact that the meaning of the quantities is not lost.    

Equation (\ref{eq:multiplicative}) represents mathematically what is called a \textit{multiplicative process}. It can be shown\cite{Cruz2014phd} that if we consider that wealth follows a multiplicative process defined like Eq. (\ref{eq:multiplicative}) then probability density function of finding an agent $x_i$ with an wealth $x$ is given by
\begin{equation}
 p(x_i=x)=\frac{\alpha x_0^{\alpha}}{x^{\alpha+1}}
 \label{eq:pareto} 
 \end{equation}
 which is the exact expression for the Paretian distribution.  Eq.(\ref{eq:pareto}) does not depend on $\beta$ as long as $\beta$ varies much more slowly than $x_i$, which is admissible since $\beta$ is an variation of the logarithm of the average of  $x_i$. 
 
 Comprehensively, there is not easy to verify empirically Eq.(\ref{eq:multiplicative}) and that is the importance of the resulting Eq.(\ref{eq:pareto}), it allow us to have a sufficient condition for the assumption taken from the definition of economic system. Also, every quantity of economic nature, like total assets\cite{Cruz2014phd}, total liabilities\cite{Cruz2014phd}, wealth\cite{Levy1997, Pareto1964} , airport traffic\cite{Albert2002} , number of internet links\cite{Albert2002}, between several other examples, follow  law that can be expressed as Eq.(\ref{eq:pareto}) with different values of $\alpha$ according to the correlation and the nature of the overall quantity. For example, total liabilities is much closed of having a zero correlation because a liability for an agent is measured as an asset in the correlated one, and $\alpha \approx 1$. On the other hand internet links are bidirectional and $\alpha \approx 2$.

Equation (\ref{eq:pareto}) gives us also the geometry of the system. If we consider an exchange as an economic link between agents, the economic system characterized by Eq.(\ref{eq:pareto}) is called a weighted scale-free network\cite{Park2004} where $x_i$ is the sum of the weights of the economic connections. This geometrical representation of an economic system allows us to state our problem: what is the multiplication effect of government spending?

\begin{defn}[Government]
\textit{Government is a channel by which all agents is a subsystem, S, of the overall economic system exchange together with some of the agents}.      
\label{def:governement}
\end{defn}

Let us take the full economic system represented in Fig.\ref{pic:scf}. The region S of the network is assumed as the relevant economic perimeter of the government and we are interested on an eventual multiplication effect of government spending on S.   From  Eq.(\ref{eq:multiplicative}) we have 
\begin{equation}
d \log \left(\frac{x_i^{\alpha}}{\Omega} \right)= 0
\label{eq:conservation}
\end{equation}   
which means that there is a quantity $E_i=\log(x_i^\alpha/\Lambda)$ that is conserved (on average, recall the definitions above) as economy evolves. From a mathematical point of view this invariance is fundamental since it establishes an ensemble: a fixed number of possible states of the system of $N$ components. Also, since $E_i$ is constant in average and $N$ are constant, then the sum of $E_i$ over the system, $E$, is also a constant.  
This is a particularly useful situation to be in because it allows us to get in the tools of statistical physics\cite{Teixeira2014}.  Since the full system amount $E$ is conserved, we can study a subsystem S.  

Now we will assume that the government of S decides to rise spending, i.e. to expend money, but keeping the same amount of product as before.  When we start speaking in money that means that $\alpha$ should be equal to  $2$ if every exchange is prefect, i.e., the amount of money in the exchange is exactly equal to the value of the product if expressed in currency units. But if the government channel exchanges money for no additional product,   that means that the correlation factor on S, $\alpha_S$, will be lower than the overall correlation factor, $\alpha$. And, knowing that $1 < \alpha < 2 $, when the spending is raised we will have
\begin{equation}
d E_S = d  \left(  \sum_S \log \left( \frac{x_i^\alpha}{\Omega} \right)  \right) \leq 0
\label{eq:diferential}
\end{equation}   

Since we did not define the size of S when compared to the economic system, S can be made of only one agent or of all of the agents. The point is that $E$ is maximized when the exchanges in the system are perfect, $\alpha=2$ and minimized when the exchanges become so unequal that each agent seems independent, $\alpha=1$. Thus, the imposition of a channel that lowers $\alpha$ in a part of a system, will lower the $E$ of the subsystem in relation with the surrounding system. Since $E$ represents the relative wealth of the subsystem, government spending is, theoretically and at the most, as good as not having government spending, i.e., free exchange between the agents. But is there no advantage on government spending? Yes, there is, on the hidden assumptions of this model that we will discuss bellow.

Another consequence of lowering $\alpha$ is on equality, one of the most spoken words in modern economics. Since $\alpha  > 1$ and $\alpha < 2$  and these values come directly from the nature of an economic system then, using Eq.(\ref{eq:pareto}) we know the maximum equality achievable in an economic system which is $\alpha=2$, i.e., when the graph of the logarithm of probability gets more vertical. That is, the same mechanisms that lower the total subsystem wealth  also lower the wealth equality.

Finally, it can be showed\cite{Cruz2015} that if we join two subsystems with different $\alpha$'s the result will be a thermalization that brings both systems to an intermediate common $\alpha$, meaning that when we have realities with more than one level of government the resulting overall wealth depends on the resulting thermalization.

 \section{Conclusions \& Discussion}
\label{sec:conclusions}

Is then government spending bad? Well, our model has some hidden assumptions that the reader should have in mind. The first and most important, and probably the one every government should take care, is that the model assumes that agents exchange freely and without constraints. That implies a considerable amount of conditions upstream provided by fundamental services, namely security (physical and social), health and education (for the existence of new things to exchange for) that cannot be provided without a channel like a government and, obviously, those imply spending. We believe our contribution is not a mean to justify cuts in or increasing government spending but, like everything in the economy, to help choices. Channeling the needs of the agents for providing free exchanges should be the role of government spending. Not only because it rises the wealth of the system but also because it improves equality.
The point here is that government spending is not virtuous \textit{per se}, in abstract if the fundamental services could be provided without spending then all government spending would be pernicious.

We can also see the problem from the other side, not having government spending at all and leaving the fundamental services to agent choices. Since that would leave agents out of free exchange system (due to lack of security or health or education), that would lower both N and wealth.

\section*{Acknowledgments}

The author likes to thank to Nuno Pessoa Barradas, Pedro Neves and Raul Vaz Os\'orio for the kind suggestions, and to Closer Consulting for the support.


 \end{document}